\documentclass[journal]{ieeetran}

\usepackage{newtxtext, newtxmath} % Times also in math mode

\usepackage{microtype}

\usepackage{booktabs}
\usepackage{cite}
\usepackage{amsmath}
\usepackage{graphicx}
\usepackage{textcomp}
\usepackage{xcolor}

\usepackage{flushend}
\usepackage[detect-all]{siunitx}
\usepackage{url}
\usepackage{dblfloatfix}
\usepackage{tablefootnote}
\usepackage{eso-pic}

\usepackage[font=footnotesize,compatibility=false]{caption}
\begin{document}

\title{Modularity is Not Enough: Demonstration of a\\Solderless 400\,V DC, 2.5\,kW Three-Phase Inverter}

\author{Luc Imperiali,~\IEEEmembership{Student Member,~IEEE,}
        Aaron Griesser,
        and~Jonas Huber,~\IEEEmembership{Senior Member,~IEEE}% <-this % stops a space
\thanks{L. Imperiali and A. Griesser are with the Advanced Mechatronic Systems Group, ETH Zurich, Switzerland, e-mail: imperiali@ams.ee.ethz.ch.}% <-this % stops a space
\thanks{J. Huber is with the Power Electronics and Drive Systems Laboratory, ETH Zurich, Switzerland.}% <-this % stops a space
}

\maketitle

% Preprint notice in the page-1 footer (DOI position), below the two-column text
\AddToShipoutPictureBG*{%
  \AtPageLowerLeft{%
    \put(\LenToUnit{0.5\paperwidth},\LenToUnit{0.8cm}){%
      \makebox[0pt][c]{%
        \begin{minipage}{0.95\textwidth}
          \centering\footnotesize
          This work has been submitted to the IEEE for possible publication.\\
          Copyright may be transferred without notice, after which this version may no longer be accessible.
        \end{minipage}%
      }%
    }%
  }%
}

\begin{abstract}
This paper presents the design and experimental evaluation of a fully solderless realization of a \qty{400}{V}, \qty{2.5}{kW} GaN-based variable speed drive (VSD), using screw-clamped resin molds and rubber compression pads instead of soldered interconnections. The power stage uses \qty{650}{V} GaN power transistors and is operated at a switching frequency of \qty{200}{kHz}. The solderless demonstrator is compared to a soldered reference realization using an identical printed circuit board (PCB). Over \num{120} thermal cycles with heatsink temperatures up to \qty{90}{\celsius}, the solderless contacts show no degradation in effective on-state resistances (including contact resistances). Separately, open-loop vibration sweeps from \qty{5}{Hz} to \qty{2}{kHz} with acceleration amplitudes above \qty{10}{g} were performed on the solderless assembly and left the continuously powered demonstrator electrically intact; subsequent resistance and nominal-power checks likewise indicate no contact degradation. An initial life-cycle assessment (LCA) indicated a higher embodied carbon footprint for the solderless realization due to 3D-printed resin molds, whereas a prospectively evaluated injection-molding scenario reduces the carbon footprint to near that of the soldered reference. The solderless assembly furthermore enables non-destructive component replacement, as demonstrated after a power transistor failure, as well as component re-use. These results support the feasibility of repair-oriented, industrially relevant kilowatt-class solderless power converters.
\end{abstract}


%%%%%%%%%%%%%%%%%%%%%%%%%%%%%%%%%%%%%%%%%%%%%%%%%%%%%%%%%%%%%%%%%%%%%%%%%%%%%%%%%%%%%%%%%%%%
\section{Introduction}
%%%%%%%%%%%%%%%%%%%%%%%%%%%%%%%%%%%%%%%%%%%%%%%%%%%%%%%%%%%%%%%%%%%%%%%%%%%%%%%%%%%%%%%%%%%%

Driven by the need to limit global warming and reduce the consumption of critical raw materials, governmental \cite{RegulationEU20242024}, industrial \cite{musilHowLifeCycle2023a}, and academic initiatives are increasingly targeting material efficiency, life cycle assessment (LCA), and ecodesign \cite{radwanLifeCycleAssessment2025,fangParametricLCAModel2025,guillemetOpenSourceLifeCycle2026,ningwangEcodesignMagneticComponents}, as well as circular economy concepts \cite{schwierzEnvironmentalLifeCycle,hellerDesignGuidelinesRepair} alongside energy efficiency \cite{huberEnergyEfficiencyNot2024}. In power electronics, this motivates a transition from purely performance-driven converter design toward systems that additionally enable repair, component reuse, and material recovery at end of life. Assessing such life-cycle aspects is becoming increasingly relevant for future converter topologies and interconnection technologies.

\begin{figure}
    \includegraphics{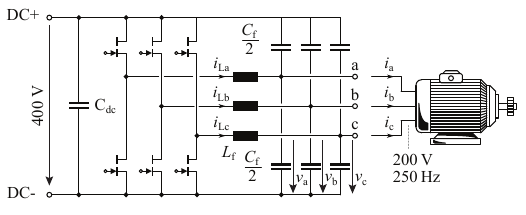}
	\caption{Power circuit of the investigated GaN-based variable speed drive (VSD) inverter featuring a dc-link-referenced LC output filter. The AC-side filter capacitance is split equally, with one half referenced to DC+ and the other half referenced to DC-, providing both differential-mode (DM) and common-mode (CM) attenuation. See \textbf{Tab.~\ref{tab:specifications}} for specifications.
    }
	\label{fig:inverter_schematics}
    \vspace{-0.5mm}
\end{figure}

Whereas megawatt and/or multi-kilovolt class power converters are often necessarily modular, and screw connections facilitate replacement of modules but also of the main components, this is normally not the case for today's ubiquitous power converters rated up to several tens of kilowatts. 
Thus, current research is investigating benefits of modularity in such systems with respect to improving compatibility with a circular economy \cite{baudaisEnvironmentalImpactModular2024, romanoCircularPowerElectronics2023}.
Depending on the definition of the modules, such design concepts would enable repair, reuse, or targeted recycling of certain functional units. 
However, virtually all power converters manufactured today rely on permanently soldered interconnections, which hinder \textit{complete} disassembly, repair, and component reuse.
Solder joints generally require destructive processes during recycling, representing a significant obstacle to circular-economy-compatible hardware. 

Several alternative interconnection approaches have been proposed to improve recoverability of all individual components mounted on a printed circuit board (PCB). Conductive adhesives can replace solder joints~\cite{conductive_paint, aljohaniElectricallyConductingWaterbased2026} but still form permanent connections that hinder component separation. Dissolvable PCBs~\cite{dissolvable_PCB} enable component recovery, but are inherently incompatible with repair scenarios.
Recently, liquid metal interconnects have been demonstrated as a reversible mounting method for power semiconductors to PCBs \cite{muRepairableRecyclableReliable2026}, but the method seems best suited for a smaller number of high-value components. 
Finally, solderless prototyping concepts essentially employing 3D-printed resin molds to press components to a PCB~\cite{sustainable_inhouse_PCB,Solderless_PCB_resin_housing,proform_thermoforming} facilitate full component recovery. However, these approaches have only been demonstrated for low-voltage signal electronics and in non-production environments. 

Solderless concepts for power electronic converters operating at several hundred volts and significant power levels in the kilowatt range remain unexplored.
Therefore, this paper demonstrates a solderless \qty{400}{V}, \qty{2.5}{kW} GaN-based variable-speed drive (VSD) system (see \textbf{Fig.~\ref{fig:inverter_schematics}}), where \textit{all} components are mounted using pressure contacts established by screw-clamped resin molds and rubber compression pads.
To the authors' knowledge, this work presents the first experimental demonstration of a fully solderless power converter at relevant power and voltage levels.
Beyond full electrical functionality, the solderless stack-up enables non-destructive component exchange and straightforward end-of-life disassembly, which are key prerequisites for repairable and recyclable converter hardware.

The remainder of this paper is organized as follows: \textbf{Section~II} presents the solderless interconnection concept, a design workflow, and the demonstrator realization. \textbf{Section~III} reports the detailed experimental characterization and benchmarking against a conventional soldered reference assembly of identical electrical specifications with respect to output-waveform quality, on-state resistance, thermal cycling behavior, vibration robustness, and embodied carbon footprint. 
\textbf{Section~IV} discusses the implications for circular power electronics and concludes the paper.

%%%%%%%%%%%%%%%%%%%%%%%%%%%%%%%%%%%%%%%%%%%%%%%%%%%%%%%%%%%%%%%%%%%%%%%%%%%%%%%%%%%%%%%%%%%%
\section{Design of a Solderless VSD Demonstrator}\label{sec:solderless_design}
%%%%%%%%%%%%%%%%%%%%%%%%%%%%%%%%%%%%%%%%%%%%%%%%%%%%%%%%%%%%%%%%%%%%%%%%%%%%%%%%%%%%%%%%%%%%

This section first introduces the solderless interconnection idea, then describes the design workflow, and finally presents the realized demonstrator with specifications summarized in \textbf{Tab.~\ref{tab:specifications}}.

\begin{table}[!t]
	\centering
	\caption{Key specifications of the variable speed drive system from \textbf{Fig.~\ref{fig:inverter_schematics}}.}
	\footnotesize
	\label{tab:specifications}
	\begin{tabular}{lcrl}
		\toprule
		Description & Symbol & Value & Unit\\
		\midrule
		Nominal motor power & $P_\mathrm{mot,N}$ & 3 & hp\\
		Nominal inverter power & $P_\mathrm{inv,N}$ & 2.5 & kW\\
		Input dc voltage & $V_\mathrm{dc}$ & 400 & V\\
		Motor voltage (l-l rms) & $V_\mathrm{M}$ & $0\dots200$ & V\\
		Motor current (rms) & $I_\mathrm{M}$ & 0\dots7.3 & A\\
		Motor el. frequency & $f_\mathrm{M}$ & $0\dots250$ & Hz\\
		\bottomrule
	\end{tabular}

\end{table}

\subsection{Solderless Interconnection Concept}
\label{ssec:solderless_concept}
\begin{figure}[!t]
	\centering
	\includegraphics[width=0.96\linewidth]{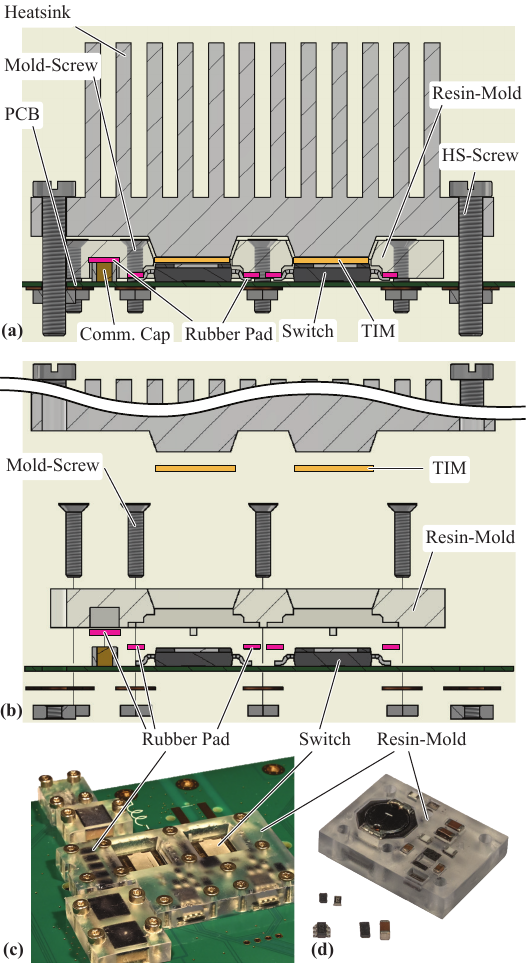}
	\caption{Mechanical stack-up of the solderless assembly concept. \textbf{(a)}~Cross-sectional view of the assembled stack showing the heatsink, power switch, thermal interface material (TIM, yellow), 3D-printed resin mold, rubber compression pad (pink), and PCB. The rubber pads and the components are placed inside the mold cavities; the PCB is placed on top and screwed down to compress the rubber pads and press the component terminals onto the PCB pads. \textbf{(b)}~Exploded view of the layer sequence and fastening screws (heatsink and mold screws). The rubber pad provides compliant compression during screw tightening, while the TIM couples the power transistors thermally to the heatsink. \textbf{(c)}~Photograph of assembled resin molds with housed components, visible through the translucent mold (heatsink not yet mounted). \textbf{(d)}~Detail view from the PCB side of a single resin mold with components placed inside.}
	\label{fig:solderless_stackup}
	\vspace{-1em}
\end{figure}

Whereas the output filter inductors (see \textbf{Fig.~\ref{fig:inverter_schematics}}) are connected via screw terminals, the following discussion applies to the mounting of all other components, including signal and control electronics, and, in particular, the power stage realized with 650\,V GaN power transistors. 
Similar to the approach proposed for low-voltage signal electronics in \cite{sustainable_inhouse_PCB,Solderless_PCB_resin_housing,proform_thermoforming}, rather than soldering the components to the PCB, components are placed inside dedicated cavities of a resin-printed mold with rubber compression pads. The PCB is placed on top, and fastening screws compress the rubber pads against the component packages to press the terminals onto the PCB pads, establishing the electrical connection without solder. The resulting mechanical stack-up is illustrated in \textbf{Fig.~\ref{fig:solderless_stackup}}.

\begin{figure*}[!b]
	\centering
	\includegraphics{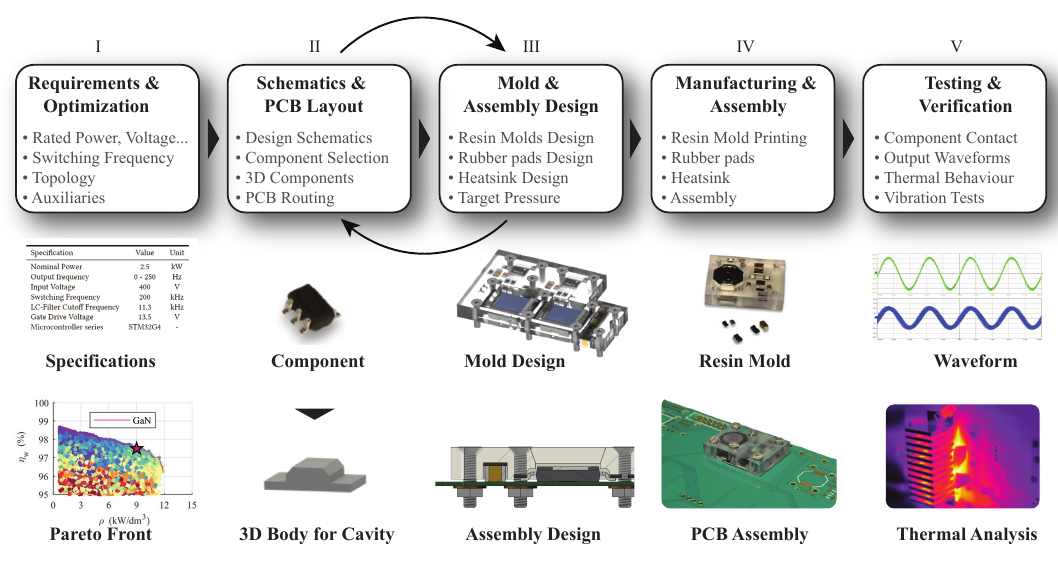}
	\vspace{0.5mm}
	\caption{Design, manufacturing, and verification workflow of the solderless VSD. \textbf{(I)}~System requirements are defined and the component set is optimized based on rated power, voltage, switching frequency, topology, and auxiliary circuitry. \textbf{(II)}~The schematic and PCB layout are developed in Altium Designer, including component selection, 3D component bodies for mold cavity generation, and power-stage routing. \textbf{(III)}~The resin molds, rubber compression pads, heatsink, and mechanical assembly are designed in a CAD software with a defined target contact pressure; steps~\textbf{(II)} and~\textbf{(III)} are iterated until electrical and mechanical constraints are met. \textbf{(IV)}~The bare PCB is fabricated, the resin molds are 3D-printed, the components are placed inside the molds, and the PCB is placed on top and screwed down. \textbf{(V)}~The assembled converter is verified through contact-resistance checks, output-waveform measurements, and thermal cycling and vibration tests.}
	\label{fig:design_workflow}
\end{figure*}

A sufficient contact force must be applied at each component terminal to ensure a stable, low-resistance connection throughout operation and thermal expansion. The rubber compression pads provide the required compliance and spring effect and have a nominal thickness of approximately \qty{0.5}{mm}. Because component heights and terminal geometries exhibit manufacturing tolerances, the pads are dimensioned for approximately \qty{0.2}{mm} compression during screw tightening, accommodating height variations while maintaining uniform contact pressure.
For smaller passives such as resistors and capacitors in standard surface-mount packages like 0603 and larger (smaller packages have not been considered so far), a single rubber compression pad is placed between the component body and the resin mold. This simple arrangement can be seen for the commutation capacitors in \textbf{Fig.~\ref{fig:solderless_stackup}(a)} and~\textbf{(b)}.

For the GaN power transistors in a top-cooled package, the mold presses only on the semiconductor leads; rubber compression pads placed at each terminal transfer the clamping force to the PCB pads rather than loading the semiconductor package, as illustrated in \textbf{Fig.~\ref{fig:solderless_stackup}(a)} and~\textbf{(b)}; other ICs like the microcontroller are handled likewise. 
The mold is furthermore designed with openings that accommodate a heatsink for the power transistors, which is placed on the switch topside with a thermal interface material (TIM) and screwed to the PCB in a second fastening step, coupling the power transistor package thermally to the heatsink without loading the electrical contacts through the mold.

Dedicated test pads on the reverse side of the PCB, adjacent to each component pad, facilitate an initial verification of the electrical connections after assembly and prior to energizing the converter, confirming correct component seating and sufficient clamping force before system testing.

Early prototype PCBs revealed an additional contact issue for very flat passives such as small resistors and capacitors: in the PCB stack-up, the top copper layer is covered by soldermask and silkscreen layers, and only the contact pads are exposed, so these low-profile components initially rested on the soldermask rather than making reliable contact to the pads. In such cases, it is thus necessary to modify the PCB footprints by removing the soldermask beneath the component bodies, ensuring direct contact between the component terminals and the PCB pads.

\subsection{Design Workflow}
\label{ssec:design_process}
With the interconnection principle established, 
the following paragraphs describe the stages~\textbf{(I)}--\textbf{(III)} of the proposed design workflow shown in \textbf{Fig.~\ref{fig:design_workflow}}, fabrication and assembly of the demonstrator are presented in \textbf{Section~\ref{ssec:demonstrator}}, and \textbf{Section~\ref{sec:results}} describes the experimental validation.

In the first stage, the requirements and converter topology are defined based on \textbf{Tab.~\ref{tab:specifications}}. A multi-objective converter optimization is performed using an optimization framework developed previously~\cite{imperialiMultiobjectiveMinimizationLifecycle2024}, and a set of components is selected from the resulting Pareto front according to the specified efficiency and power-density targets. Note that any other design method could be employed, in principle.

The second stage covers the detailed PCB design in Altium Designer. Schematics are drafted and specific components are selected for the power stage, gate-driver circuitry, and auxiliary functions. For each component later housed inside a resin mold, a simplified three-dimensional body is defined on a separate layer, representing the occupied volume for export to the mechanical design in stage~\textbf{(III)}. Further, the soldermasks of certain component footprints must be adjusted to ensure proper contact as discussed above. Once the schematics are complete, component placement and routing are carried out in the PCB editor. 

The third stage comprises the mechanical design in a CAD software such as Autodesk Inventor. The finalized PCB layout is imported into the CAD environment, where the resin molds, heatsink, and screw-fastening features are modeled. The abstract 3D bodies from Altium Designer are subtracted from the mold solids to obtain cavities with the correct shape and position for each component. Because this Boolean subtraction is straightforward to update, steps~\textbf{(II)} and~\textbf{(III)} of \textbf{Fig.~\ref{fig:design_workflow}} can be executed iteratively until PCB pad dimensions, mold envelopes, screw locations, and clearances are simultaneously satisfied. The dimensions of the rubber compression pads are derived directly from the generated mold geometry.

\subsection{Realization}
\label{ssec:demonstrator}
\begin{figure}[t]
	\includegraphics[width=\linewidth]{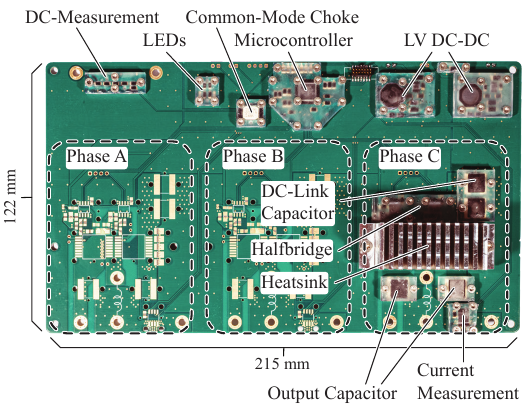}

	\caption{Top view of the realized solderless VSD demonstrator board with outer dimensions of \qty{215}{mm}~$\times$~\qty{122}{mm}. The board comprises three phase legs (\textbf{A}--\textbf{C}) and shared control/auxiliary circuitry along the upper edge. Phase~\textbf{C} is fully assembled with resin 3D-printed molds and heatsink; phases~\textbf{A} and~\textbf{B} show the bare PCB pad layout. For simplicity, only Phase~\textbf{C} was assembled for this first prototype.}
	\label{fig:solderless_pcb}
	\vspace{0.5mm}
\end{figure}

\textbf{Fig.~\ref{fig:solderless_pcb}} shows the realized demonstrator that integrates three identical phase legs based on Infineon IGLT65R045D2 GaN power transistors (only one phase leg was assembled for demonstration purposes).
Each phase leg comprises a half-bridge with gate drivers, output capacitors, and current measurement, together with mounting holes for the heatsink and resin molds. Along the upper edge, the shared control and auxiliary circuitry includes an STM32G474 microcontroller, auxiliary power supplies using low-voltage dc-dc converters, a dc-link voltage measurement including a signal common-mode choke, and status LEDs. 

For this prototype, separate molds were used for each functional unit to simplify debugging and mechanical iteration, at the cost of higher initial assembly effort. This reflects a design trade-off between selective repairability, which enables individual component groups to be exchanged without full disassembly, and manufacturing complexity, which can be optimized in future designs by consolidating molds, e.g., to one mold per phase leg or for the entire power stage.

Phase~\textbf{C} was assembled according to workflow stage~\textbf{(IV)}. The resin molds were 3D-printed from clear resin~\cite{clearresin_formlabs} using masked stereolithography (MSLA), and the rubber compression pads were made with a CNC laser cutter. The power semiconductors and associated passives were placed inside the mold cavities, and the PCB was placed on top and screwed down to the intended terminal pressure. Contact integrity was checked on the reverse-side test pads before installing the TIM and heatsink. The electrical and thermal verification of stage~\textbf{(V)} is reported in \textbf{Section~\ref{sec:results}}.

%%%%%%%%%%%%%%%%%%%%%%%%%%%%%%%%%%%%%%%%%%%%%%%%%%%%%%%%%%%%%%%%%%%%%%%%%%%%%%%%%%%%%%%%%%%%
\section{Results}\label{sec:results}
%%%%%%%%%%%%%%%%%%%%%%%%%%%%%%%%%%%%%%%%%%%%%%%%%%%%%%%%%%%%%%%%%%%%%%%%%%%%%%%%%%%%%%%%%%%%

The solderless demonstrator is compared to a conventional soldered reference board of identical layout, with only minor adjustments to the soldermask layer (as discussed above). Characterization was performed on a single phase leg at a dc-link voltage of \qty{400}{V} and a switching frequency of \qty{200}{kHz}. Nominal-power tests employ an external \qty{100}{\micro\henry} output inductor and an adjustable load resistor connected between the phase output and the midpoint of an external dc-link capacitor bank, yielding approximately \qty{830}{W} per phase and an RMS phase current of \qty{7.3}{A}. The following subsections report the electrical performance and thermal cycling behavior relative to the soldered reference, a vibration campaign performed on the solderless PCB, and a comparison of the embodied carbon footprints.

\subsection{Electrical Performance}
\label{ssec:electrical_performance}

Commissioning began with zero-current-switching (ZCS) tests at dc-link voltages up to \qty{400}{V}. For both assemblies (soldered and solderless), the switch-node voltage exhibited comparable slopes exceeding \qty{100}{V/\nano\second}, confirming that the solderless contacts do not impair the switching transition.

\begin{figure}[t]
	\includegraphics{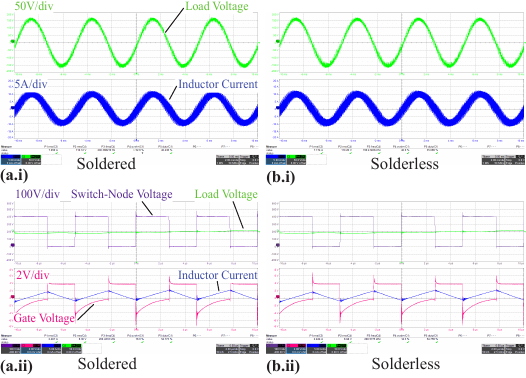}
	\caption{Measured waveforms of the soldered reference \textbf{(a)} and the solderless demonstrator \textbf{(b)} at nominal single-phase power (\qty{400}{V} dc-link voltage, \qty{\sim 830}{W}). \textbf{(a.i)} and~\textbf{(b.i)} show the filtered load voltage in green and the inductor current in blue over four output cycles. \textbf{(a.ii)} and~\textbf{(b.ii)} resolve the switching interval at \qty{5}{\micro\second/div}, displaying the switch-node voltage in violet, the load voltage in green, the low-side gate voltage in red, and the inductor current in blue.}
	\label{fig:waveforms}
\end{figure}

Under loaded conditions at nominal single-phase power, both realizations produce the same sinusoidal load voltage and inductor current over four fundamental output cycles, and a zoomed view of the switching interval reveals effectively indistinguishable key waveforms (see \textbf{Fig.~\ref{fig:waveforms}}). These results are consistent with the preceding ZCS tests and indicate that the solderless interconnections do not prevent correct inverter operation at nominal power.

\subsection{Thermal Cycling Tests}
\label{ssec:thermal_cycling}

\begin{figure}[t]
	\includegraphics{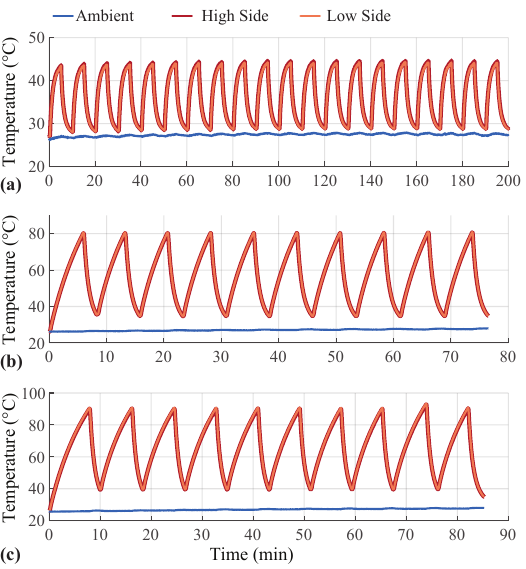}
	\caption{Recorded heatsink temperatures during thermal cycling of the solderless demonstrator. \textbf{(a)}~Example of \num{20} automated cycles with \qty{5}{min} of heating and \qty{5}{min} of cooldown ($\Delta T \approx \qty{20}{\celsius}$); five such campaigns were performed for cycles~\num{1}--\num{100}. \textbf{(b)}~Cycles~\num{101}--\num{110} with an enlarged heatsink temperature swing between \qty{35}{\celsius} and \qty{80}{\celsius} ($\Delta T \approx \qty{45}{\celsius}$). \textbf{(c)}~Cycles~\num{111}--\num{120} with a further enlarged temperature swing between \qty{40}{\celsius} and \qty{90}{\celsius} ($\Delta T \approx \qty{50}{\celsius}$). High-side and low-side heatsink temperatures are shown in red and orange; ambient temperature is shown in blue.}
	\label{fig:thermal_cycles}
\end{figure}

Having established comparable switching and output behavior, the contact stability of the solderless assembly was evaluated under repeated thermo-mechanical loading and compared to the soldered reference converter.
Both converters were subjected to automated power cycling at nominal single-phase power: \qty{5}{min} of full-load operation followed by \qty{5}{min} of cooldown, repeated for \num{100} cycles. During each cycle, the heatsink temperature swings from approximately \qty{25}{\celsius} to \qty{45}{\celsius} ($\Delta T \approx \qty{20}{\celsius}$), as illustrated for a representative sequence in \textbf{Fig.~\ref{fig:thermal_cycles}(a)}. Heatsink temperatures were logged throughout each cycle using sensors inserted into drilled holes above the high-side and low-side semiconductors. 

\begin{figure}
	\includegraphics{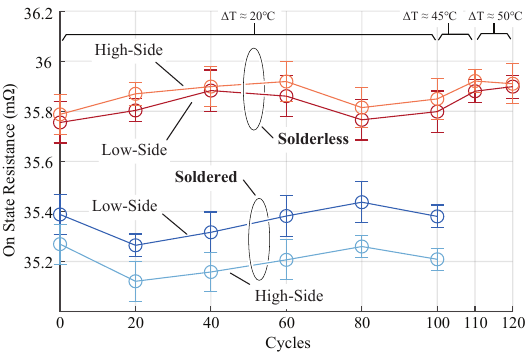}
	\caption{Temperature-compensated effective on-state resistance as a function of the thermal cycle count for the soldered reference and the solderless VSD demonstrator. The plotted quantity includes both the intrinsic $R_{\mathrm{DS,on}}$ and the PCB contact resistance, because the voltage drop is sensed close to the PCB pads. Error bars indicate the Keysight~34410A multimeter measurement uncertainty. During cycles~\num{0}--\num{100}, both converters were power-cycled with \qty{5}{min} of loaded operation and \qty{5}{min} of cooldown ($\Delta T \approx \qty{20}{\celsius}$), with the resistance measured after every \num{20} cycles. On the solderless board only, ten additional cycles with $\Delta T \approx \qty{45}{\celsius}$ (heated to \qty{80}{\celsius}, cooled below \qty{35}{\celsius}) and ten further cycles with $\Delta T \approx \qty{50}{\celsius}$ (heated to \qty{90}{\celsius}, cooled below \qty{40}{\celsius}) were performed, with measurements at cycles~\num{110} and~\num{120}. Blue colors refer to the soldered reference and red colors to the solderless realization. No relevant change is observed over the full sequence, indicating stable pressure contacts under increasing thermal stress.}
	\label{fig:rdson}
\end{figure}

After every \num{20} cycles, the effective on-state resistance of both switches was measured on the deenergized converter (without any hardware modifications made, in particular without removing the heatsink or the molds). The resistance was determined from a 1\,A drain-source current and from the voltage drop sensed close to the PCB pads, so that the measured value includes both the intrinsic drain-source resistance $R_{\mathrm{DS,on}}$ and the contact resistance to the PCB. Voltage and current were acquired with Keysight~34410A precision multimeters; the corresponding instrument uncertainties are shown as error bars in \textbf{Fig.~\ref{fig:rdson}}. A gate-drive leakage current of approximately \qty{20}{mA} through the permanently biased high-side gate was compensated in the low-side measurement. Measured values were also temperature-compensated to a reference temperature of \qty{25}{\celsius} using the temperature coefficient from the device datasheet~\cite{Infineon_semi}.

To further stress the solderless assembly, two additional ten-cycle tests with enlarged temperature swings were performed on the solderless board only. In the first campaign, the fans were disabled and the converter was operated at full power until both heatsink sensors reached \qty{80}{\celsius}; the converter was then shut off, the fans were re-enabled, and cooldown continued until the heatsink temperature fell below \qty{35}{\celsius} ($\Delta T \approx \qty{45}{\celsius}$). This sequence was repeated ten times, and $R_{\mathrm{DS,on}}$ was measured at cycle~\num{110} (see \textbf{Fig.~\ref{fig:thermal_cycles}(b)}). In the second campaign, the same procedure was applied with a peak temperature of \qty{90}{\celsius} and a cooldown target below \qty{40}{\celsius} ($\Delta T \approx \qty{50}{\celsius}$), again for ten cycles (\textbf{Fig.~\ref{fig:thermal_cycles}(c)}), followed by a measurement at cycle~\num{120}.

\textbf{Fig.~\ref{fig:rdson}} summarizes the effective on-state resistance measurement results for both converters over cycles~\num{0}--\num{100} and for the solderless demonstrator through cycle~\num{120}. Immediately after assembly, the temperature-compensated on-resistance of the solderless switches was approximately \qty{36}{m\ohm}, compared to roughly \qty{35.5}{m\ohm} for the soldered reference. Both values lie below the datasheet typical value of \qty{45}{m\ohm} at \qty{25}{\celsius}, most likely because the switches are driven with a higher gate-drive current than the \qty{33}{mA} used for datasheet characterization~\cite{Infineon_semi}. More importantly, the \qty{0.5}{m\ohm} difference between assemblies is small relative to the manufacturing tolerance between the typical and maximum ratings (\qty{45}{m\ohm}--\qty{54}{m\ohm}), so no systematic penalty for the solderless assembly can be inferred.
At the nominal RMS phase current of \qty{7.3}{A}, this difference increases conduction losses by only about \qty{26}{mW} per switch, which is negligible relative to the \qty{830}{W} output power.

No relevant systematic changes in the effective on-state resistances of the solderless assembly are observed over the full test sequence. The minor fluctuations are similar for the solderless realization and for the soldered baseline, mostly within or close to the measurement device accuracy, and the remainder is likely attributable to some uncontrolled variables like ambient temperature, etc.
This initial thermal cycling test sequence indicates that the solderless pressure contacts remain stable under repeated and significant temperature changes.

\begin{figure}[t]
	\includegraphics{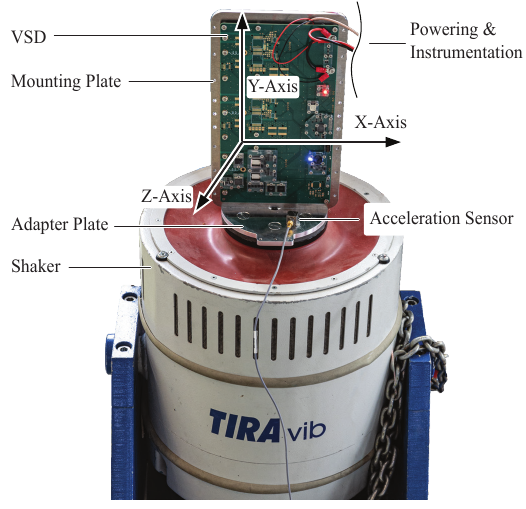}
	\caption{Vibration test setup with the solderless VSD demonstrator mounted in Y-direction on a TIRA Vib 5520-120 electrodynamic shaker. The PCB is fastened to a mounting plate that is bolted via an adapter plate to the shaker interface. An acceleration sensor on the adapter plate records the acceleration experienced by the assembly. Throughout the campaign, the board remains continuously powered from a \qty{12}{V} auxiliary supply and a reduced dc-link voltage of \qty{12}{V}, while the phase-leg switch-node voltage is monitored on an oscilloscope; the corresponding cables therefore remain connected so that any contact failure can be detected immediately. The indicated $X$, $Y$, and $Z$ axes define the three shaking directions (the PCB mounting orientation is changed w.r.t. the shaker interface accordingly). The shaker is operated in open loop.}
	\label{fig:vibration_setup}
\end{figure}

\subsection{Vibration Tests}
\label{ssec:vibration}

To assess whether the pressure contacts of the solderless assembly remain intact under mechanical stress representative of industrial environments, the solderless demonstrator was subjected to a vibration campaign on a TIRA Vib 5520-120 electrodynamic shaker. Prior to these tests, all mold and fixture screws were removed and reinstalled with a thread-locking adhesive (Loctite 243). 
Further, to focus on the presented solderless assembly technique, the relatively large and non-optimized heatsink was not installed for the vibration tests.
After this complete reassembly, the converter operated correctly on the first attempt, providing a further confirmation that the pressure contacts can be opened and restored (i.e., the mold and components removed and remounted) reliably without soldering.

\textbf{Fig.~\ref{fig:vibration_setup}} shows the instrumented vibration test setup. The board was kept continuously powered throughout the tests, so that a contact interruption would appear immediately in the switch-node waveform. For reduced risk, the dc-link voltage was limited to \qty{12}{V} rather than the nominal \qty{400}{V}. Because the shaker can be operated only in open loop, several familiarization sweeps were first performed to establish suitable drive settings. The subsequent campaign targeted accelerations above \qty{10}{g} over a frequency range from \qty{5}{Hz} to \qty{2}{kHz} along the three axes, exceeding typical severity levels of sinusoidal vibration tests according to IEC~60068-2-6~\cite{IEC60068-2-6}.

\begin{figure}
	\includegraphics{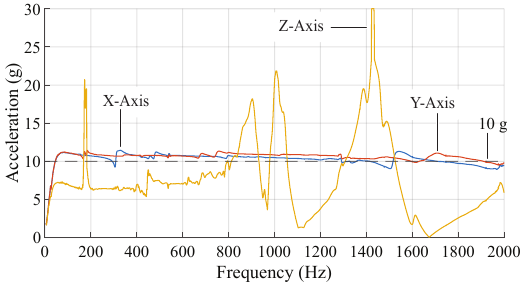}
	\caption{Measured acceleration profiles over a frequency sweep from \qty{5}{Hz} to \qty{2}{kHz} along the three axes indicated in \textbf{Fig.~\ref{fig:vibration_setup}}; the accelerometer was mounted on the adapter plate. The dashed line marks a \qty{10}{g} reference. Because the shaker is operated in open loop, the excitation level is not held constant across the sweep; the measured acceleration instead reveals the structural response including resonant modes. Along $X$ and $Y$, the response remains comparatively flat. In the $Z$ direction, with the board mounted horizontally on the shaker, several resonant modes appear, with pronounced peaks near \qty{170}{Hz}, \qtyrange{800}{1000}{Hz}, and around \qty{1.45}{kHz}.}
	\label{fig:vibration_test}
\end{figure}

\textbf{Fig.~\ref{fig:vibration_test}} reports the measured accelerations for the three shaking directions. Along the $X$- and $Y$-directions, the response stays relatively flat over most of the considered frequency range, whereas the $Z$-axis orientation, with the board mounted horizontally on the shaker, exhibits several resonant peaks well above \qty{10}{g}. 

Under these frequency/acceleration sweeps, the demonstrator continued to operate without interruption of the switch-node waveform, indicating that the solderless assembly interconnections remained electrically intact.
Following the vibration campaign, the effective on-state resistance, again comprising $R_{\mathrm{DS,on}}$ and the PCB contact resistance, of the power transistors was remeasured and found to be within the ranges observed in \textbf{Fig.~\ref{fig:rdson}}. The board was subsequently operated at nominal power (\qty{400}{V} dc, approximately \qty{830}{W} single-phase output power) without functional issues. A further resistance measurement after this loaded test likewise indicated no degradation of the pressure contacts.

\subsection{Embodied Carbon Footprint}
\label{ssec:co2e}

A comparison of the embodied carbon footprint (CO$_2$eq) was performed for a soldered reference design, the presented solderless VSD demonstrator, and a second solderless realization scenario with injection-molded housings instead of 3D-printed resin molds. The functional unit is the complete three-phase converter PCB, including the power stage and all auxiliary circuitry. The inventory data was compiled following the LCA framework established in our previous work, i.e., the component-level embodied CO$_2$eq of the auxiliary components, semiconductors, heatsink, and capacitors was estimated as in~\cite{imperialiMultiobjectiveMinimizationLifecycle2024} and complemented by data from the ecoinvent database~\cite{ecoinventEcoinventDatabase}, e.g., for resin 3D printing, etc.

For the soldered reference, the PCB area of a volume-optimized \qty{2.5}{kW} VSD design with identical electrical specifications was used~\cite{Littledrive}, yielding \qty{43.1}{\square\centi\meter}. As shown in \textbf{Fig.~\ref{fig:solderless_pcb}}, the realized solderless demonstrator occupies a disproportionately large PCB area, as it represents a first prototype with separate molds per component group for debugging and mechanical iteration. For the solderless inventory, the total PCB area was obtained by adding the footprint area of all resin molds, which is \qty{95.4}{\square\centi\meter}, i.e., the required PCB area is about two times that of the soldered reference design.

The presented solderless realization additionally includes the 3D-printed resin molds, fastening screws, and rubber compression pads. The embodied CO$_2$eq of the resin molds was calculated from their measured mass and a resin-printing process inventory~\cite{resin_printing}. The rubber pads were modeled from the total mass, and include laser cutting as the manufacturing process. Considering a future industrial scenario, the same mold mass was retained, but the manufacturing process was changed from resin printing to injection molding with ABS thermoplastic polymer using ecoinvent data~\cite{ecoinventEcoinventDatabase}. 

\begin{figure}[t]
	\includegraphics[width=\linewidth]{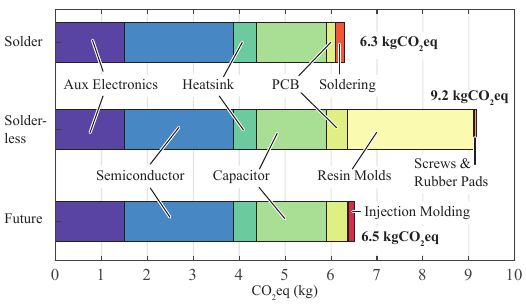}
	\caption{Embodied carbon footprint (CO$_2$eq) of the soldered reference design, the presented solderless demonstrator with 3D-printed resin molds, and a future solderless realization with injection-molded housings, broken down by component group. The auxiliaries, semiconductors, heatsink, and capacitors contribute equally to all three realizations, whereas the PCB contribution reflects a volume-optimized soldered reference (\qty{43.1}{\square\centi\meter}) and the larger solderless prototype footprint (\qty{95.4}{\square\centi\meter}) to accommodate the molds, respectively. The presented solderless assembly replaces the soldering step with 3D-printed resin molds as well as screws and rubber pads; in the future scenario, the resin molds are replaced by injection-molded parts with a substantially lower embodied carbon footprint.}
	\label{fig:co2e_comparison}
\end{figure}

\textbf{Fig.~\ref{fig:co2e_comparison}} presents the resulting embodied carbon footprint breakdowns and clearly shows that the electronic components dominate all three scenarios. For the soldered reference, the total embodied footprint is approximately \qty{6.3}{kg}~CO$_2$eq, including the soldering step. The presented solderless realization adds approximately \qty{2.9}{kg}~CO$_2$eq, driven primarily by the 3D-printed resin molds, yielding a total of approximately \qty{9.2}{kg}~CO$_2$eq. In a future industrial scenario with injection-molded ABS housings instead of resin-printed molds, the total embodied carbon footprint decreases to approximately \qty{6.5}{kg}~CO$_2$eq, close to that of the soldered reference VSD.

%%%%%%%%%%%%%%%%%%%%%%%%%%%%%%%%%%%%%%%%%%%%%%%%%%%%%%%%%%%%%%%%%%%%%%%%%%%%%%%%%%%%%%%%%%%%
\section{Conclusion and Outlook}\label{sec:discussion}
%%%%%%%%%%%%%%%%%%%%%%%%%%%%%%%%%%%%%%%%%%%%%%%%%%%%%%%%%%%%%%%%%%%%%%%%%%%%%%%%%%%%%%%%%%%%
This work demonstrates that a completely solderless assembly of a kilowatt-class power converter is feasible, using screw-clamped resin molds and rubber compression pads instead of soldered component interconnections. Against a soldered reference PCB of identical layout, the demonstrator shows comparable switch-node waveforms and stable effective on-state resistances (including contact resistances) over \num{120} thermal cycles. Separately, vibration sweeps from \qty{5}{Hz} to \qty{2}{kHz} with accelerations exceeding \qty{10}{g} in all three axes left the continuously powered PCB electrically intact. Subsequent resistance and nominal-power checks show no contact degradation. Together, these results indicate that the pressure contacts remain stable under both thermo-mechanical and vibration loading in the tested range.

Whereas the prototype realization of the molds using resin 3D-printing comes with a clearly higher embodied carbon footprint compared to a soldered reference design, this drawback diminishes when considering an industrial scenario with injection-molding instead. This indicates that the manufacturing-phase penalty observed here is largely governed by the mold fabrication technology rather than by the reversible interconnection principle itself.
From a circular-economy perspective, however, the solderless approach offers clear advantages: components housed in removable molds can be exchanged non-destructively and separated at end of life. During a pretest, a semiconductor overheated above \qty{100}{\celsius} and had to be replaced; the repair was completed quickly, whereas a soldered assembly would have required desoldering with a higher risk of collateral damage.

Several mechanical design vectors remain open. Mold envelopes should be size-optimized toward a production-like footprint, and screw locations should be co-optimized with electrical layout and contact pressure. Alternative fastening methods that reduce or eliminate screws, for example clamp tabs or spring clips, may further lower assembly effort while preserving reversible access. Closely related is manufacturability: the presented prototype still requires substantial manual assembly, and future work should quantify this effort and identify which steps, cavity generation from PCB component bodies, pad cutting, mold printing or injection molding, and component placement by pick-and-place, can be automated.

Reliability qualification should be extended beyond the presented initial thermal-cycling test and open-loop vibration campaign to include humidity, shock, and drop tests, together with dedicated corrosion studies of the pressure contacts. Selective plating of component terminations could improve contact stability, but any such coating must be assessed against the additional embodied carbon and other environmental impacts it introduces.

The life-cycle comparison in this work is limited to manufacturing-phase (embodied carbon footprint,  CO$_2$eq). A fuller assessment should include further impact categories such as resource depletion, toxicity, and water use, as well as use-phase scenarios including repair loops, mold and component reuse, and end-of-life recovery of components and materials.

In summary, the presented solderless assembly concept extends reversible interconnection to a \qty{400}{V}, \qty{2.5}{kW} GaN VSD demonstrator without compromising electrical, thermal, or vibration performance in the tested operating range. With mold fabrication adapted beyond prototyping resin printing, the approach offers a viable route toward repairable and recyclable power converters.

%%%%%%%%%%%%%%%%%%%%%%%%%%%%%%%%%%%%%%%%%%%%%%%%%%%%%%%%%%%%%%%%%%%%%%%%%%%
\section*{Acknowledgment}
%%%%%%%%%%%%%%%%%%%%%%%%%%%%%%%%%%%%%%%%%%%%%%%%%%%%%%%%%%%%%%%%%%%%%%%%%%%
The authors would like to thank the \textit{European Center for Power Electronics e.V. (ECPE)} for the financial support and Dominik Werne and the Institute of Structural Mechanics and Monitoring (IBK) at ETH Zurich for supporting the vibration tests. The Advanced Mechatronic Systems Group at ETH Zurich is generously supported by the \textit{Else und Friedrich Hugel Fonds} via the \textit{ETH Zurich Foundation}, for which the authors are most grateful.

%%%%%%%%%%%%%%%%%%%%%%%%%%%%%%%%%%%%%%%%%%%%%%%%%%%%%%%%%%%%%%%%%%%%%%%%%%%%%%%
% \balance
\linespread{1}
\bibliographystyle{IEEEtran}
\bibliography{bibliography}
	
\end{document}